# Optimization of Production Protocol of Alkaline Protease by Streptomyces pulvereceus

[1]D.Jayasree, [2]T.D. Sandhya Kumari, [3]P.B.Kavi Kishor, [4]M.Vijaya Lakshmi, [5]M. Lakshmi Narasu
[1]Centre for Biotechnology, Institute of Science and Technology,
Jawaharlal Nehru Technological University Hyderabad,
Kukatpally, Hyderabad – 500085 Andhra Pradesh, India.
jayasreekiran@yahoomail.co.in
[2]Department of Genetics, Osmania University, Hyderabad – 500 007
[3]Department of Botany, Acharya Nagarjuna University, Nagarjuna Nagar -522 507

## Abstract

**Extra cellular alkaline protease producing species is an isolate from soil which was characterized and identified as *Streptomyces pulvereceus* MTCC 8374. Studies on submerged fermentation revealed that maximum level of enzyme production was during early stationary phase. Optimum pH, inoculum and temperature were 9.0, 3 percent and 33 C respectively. Among carbon sources 0.3 peccent starch gave a maximum production followed by maltose, xylose and fructose. High yield of protease production was reported with 1.0 percent casein followed by soybean meal, yeast extract and malt extract. Further, it was optimized with 0.5 percent, 1.0 percentand 1.5 percent of NaCl among which 1 percent NaCl resulted in maximum level of protease. The protease profile of the isolate shows its potential as a good source for industrial application.**

## 1. INTRODUCTION

Proteases are the most important group of industrial enzymes which account for about 60% of total enzymes in the market. Bacterial proteases are most significant compared with animal and fungal proteases and have wide range of industrial application. Among different proteases, alkaline proteases produced by microorganisms are of main interest from a biotechnological perspective, and are investigated not only in scientific fields of protein chemistry and protein engineering but also in applied fields such as detergents, foods, tannery, pharmaceutical, and leather industries. A protease catalyses the hydrolysis of proteins into peptides and amino acids, and consists of one of the most useful enzyme groups.

The possibility of using *Streptomyces* for protease production has been investigated because of their capacity to secrete the proteins into extra cellular media, which is generally regarded as safe with food and drug administration (GRAS). *Streptomyces sps* that produce proteases include *S. clavuligerus, S. griseus, S. rimouses, S. thermoviolaceus*, *S. thermovulgaris* (James et al 1991; Edward et al., 1994). *Streptomyces sps* are heterotrophic feeders which can utilize both complex and simple molecules as nutrients. In addition to antibiotics *Streptomyces species* liberate several extra cellular enzymes (Gupta et al.1995). They produce variety of extra cellular proteases that have been related to aerial mycelium formation and sporulation (Kim and Lee 1995).

Most of the alkaline proteases applied for industrial purpose have limitations, first many of the alkaline proteases exhibit low activity and low stability at wide range of pH and temperature and secondly 30-40% of production cost of industrial enzymes is estimated to be accounting for the cost of growth medium. As the composition of culture medium strongly influences enzyme production (Giarrhizzo et al. 2007) an attempt was made to optimize cultural conditions and select an isolate with higher tolerance against a wide range of pH and temperature for getting higher yield of protease enzyme.

## 2. MATERIALS AND METHODS

**Sample collection:** Samples were collected from various places in and around Warangal, A.P, India. Soil samples were collected from the dumpyards of Food Corporation of India, Bread factories and Dairy farms.

**Isolation and screening of organisms** The culture was serially diluted and 0.1 ml of culture was plated on screening medium for isolation of pure cultures.

**Screening for proteolytic activity**

Proteolytic activity from isolated pure cultures was screened by plating on Casein agar [Casein 0.3g, $KNO_3$ 0.2g, NaCl 0.2g, $K_2HPO_4$ 0.2g, $MgSO_4$ 0.005g, $CaCl_2$ 0.002g, Yeast extract, 0.1g; agar 2g, distilled water 100ml].The organism having maximum caseinolytic activity was selected and maintained on Luria Bertani agar (LB) slants at 4 ºC.

**Characterization**: The culture was identified based on the morphological and biochemical characteristics and same was sent to Institute of Microbial Technology (IMTECH), Chandigarh, for authentic identification.

**Nutritional factors affecting growth and protease production.** The entire study on nutritional factors affecting the growth and protease production was performed as follows: 1.0 ml of 24 h bacterial culture was inoculated in to 100 ml of medium and incubated at 30 °C for 168 h. Basal medium (containing glucose 0.5 g/l; peptone, 10 g/l; $KNO_3$, 0.6 g/l; NaCl, 5g/l; $K_2HPO_4$, 0.5 g/l; $MgSO_4$. 7 $H_2O$, 0.5 g/l; $CaCl_2$, 1.0 g/l) was used for inoculation. The culture was centrifuged at 10,000 rpm for 10 min at 4°C to separate the clear supernatant. Samples were withdrawn at 24h intervals to check protease production and to determine biomass. The supernatant was analyzed for protease activity by using casein as substrate (Hagihara *et. al.* 1958) and biomass was calculated by wet weight (g/l).

**Effect of pH, inoculum, and temperature on protease production:** Effect of pH on growth and protease production was studied by adjusting pH of basal media ranging from 7-11. Effect of inoculum on protease production was studied by incubating the medium with inoculum ranging from 1-5 percent. Influence of temperature on growth and protease production was studied by incubating the culture media at temperatures ranging from 27°C- 40°C. After 48 h of incubation, the culture was centrifuged at 10,000 rpm for 10 min at 4°C to separate the clear supernatant. The supernatant was analyzed for protease activity by casein assay.

**Effect of carbon sources on protease production:** To test the effect of different carbon sources on protease production, glucose in the basal medium was substituted with (0.3%w/v) of starch, fructose, xylose and maltose. All carbon sources were filter sterilized by 0.22 µm (Millipore, USA) membrane filters. After sterilization of media sugar was added and organism was inoculated and incubated for 48 h at 33 °C, 180 rpm. This organism was checked for maximal caesinolytic activity.

**Effect of organic nitrogen source on protease production:** To test the effect of different organic nitrogen sources on protease production, peptone in the basal medium was substituted with 1% w/v of various organic nitrogen sources; malt extract, casein, soy meal extract and yeast extract. After sterilization, the organism was inoculated and incubated for 48 h at 33 °C, in an orbital shaker (180 rpm). This isolate was checked for its caesinolytic activity.

**Effect of NaCl on protease production:** To study the effect of NaCl on protease production, isolate was cultured in optimized basal medium, containing 0.5%, 1.0%, 1.5% NaCl. The cultures were incubated at 33°C for 48 h and analyzed for growth and protease activity.

## 3. RESULTS AND DISCUSSION

**Isolation and Screening for protease producing organisms** Isolates were initially screened based on zone of clearance on agar plates shown in (**Table 1**). Among the 10 isolates ‘S1’ was found to have higher protease activity and was selected and further studied.

Table 1

| Isolate | Zone of Hydrolysis in ( mm) |
|---|---|
| K1 | 18 |
| K2 | 22 |
| K3 | 24 |
| M1 | 26 |
| M2 | 16 |
| M3 | 22 |
| P1 | 27 |
| P2 | 29 |
| P3 | 19 |
| S1 | 32 |

**Identification of isolate:** Morphological and biochemical characteristics of the isolate S-1 were studied and compared with the standard characteristics described in Bergy’s manual. It is a gram positive bacterium with white coloured colonies. Based on morphology and biochemical characterization, given in (Table 2) the isolate was identified as *Streptomyces pulveraceus* MTCC 8374.

Table 2 Characters of the organism

| TESTS | S-1 |
|---|---|
| Gram's staining | + |
| Endo spore | - |
| Acid fast staining | - |
| Cell morphology | Mycelial |
| Citrate utilization | - |
| Starch hydrolysis | + |
| Casein hydrolysis | + |
| Phenyl alanine test | - |
| Hydrogen sulphide production | - |
| Nitrate reduction | - |
| Catalase | + |
| Oxidase | - |
| Growth on 2% NaCl | + |
| 5% NaCl | + |
| 7% NaCl | + |
| 10% NaCl | + |

The protease production of *Streptomyces pulveraceus* started at 24 h and reached to the maximum level after 72 h of cultivation. There was a gradual increase in biomass as well as protease production during stationary phase. These results confirmed the observation of (Petinate et al., 1999; Yang and Wang 1999) who reported an increase of protease production by *Streptomyces cyanens* and *S. rimosus* during log phase. This indicates that high level of protease production is observed during active biomass production. Effect of incubation time on growth and protease production by *Streptomyces pulveraceus* is shown in (**Table 3**)

Table 3 Growth and activity of protease

| Time (h) | Biomass (g/L) | Protease activity(U) |
|---|---|---|
| 24 | 0.632 | 30 |
| 48 | 1.02 | 86 |
| 72 | 1.16 | 186 |
| 96 | 1.52 | 252 |
| 120 | 1.12 | 90 |
| 144 | 0.92 | 60 |
| 168 | 0.72 | 5 |

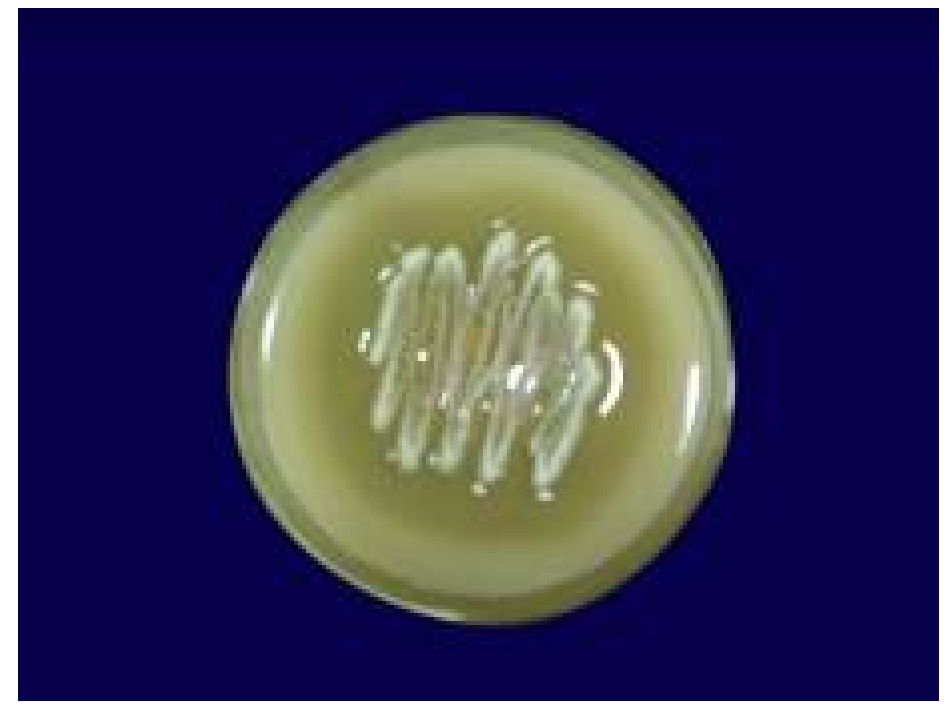

Figure 1 Casein hydrolysis of the isolate

The impact of pH, temperature, inoculum, carbon, nitrogen sources and sodium chloride on protease production was studied. The optimum conditions for the enzyme production in *Streptomyces pulveraceus* are pH 9.0, inoculum 3%, temperature 33ºC among parameters tested (Fig. 2). Starch triggered maximum protease production followed by maltose, fructose and xylose (Fig. 3). Among nitrogen sources, casein affected maximum protease activity followed by soybean meal, yeast extract, and malt extract (Fig. 4). The highest yield of protease production was noticed with 0.3% starch as a sole carbon source and 1% casein as a sole nitrogen source and 1% NaCl. (Fig: 5)

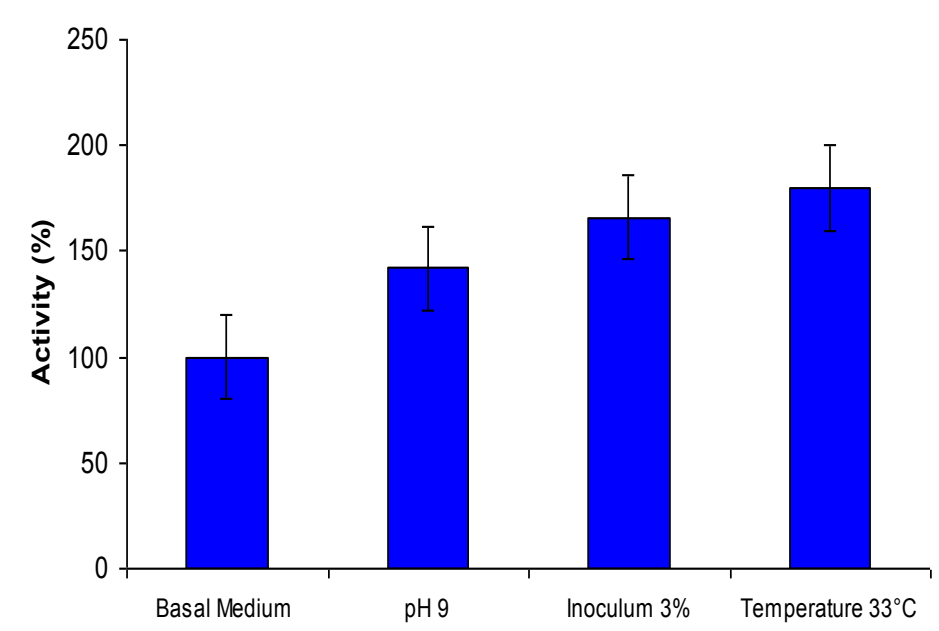

Figure 2 Bacterial growth & protease production at optimum pH, inoculum & temperature

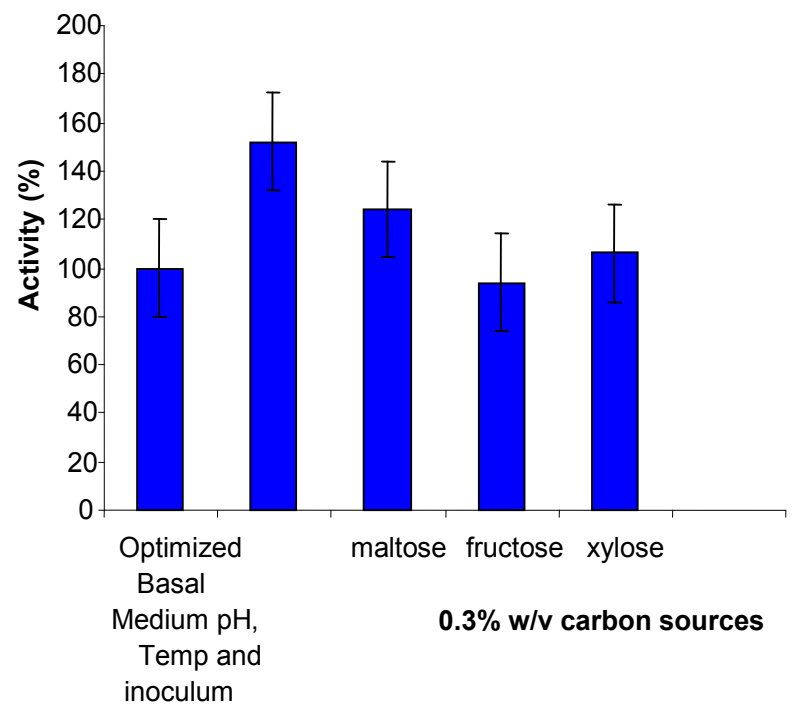

Figure 3 Effect of Carbon sources on protease production

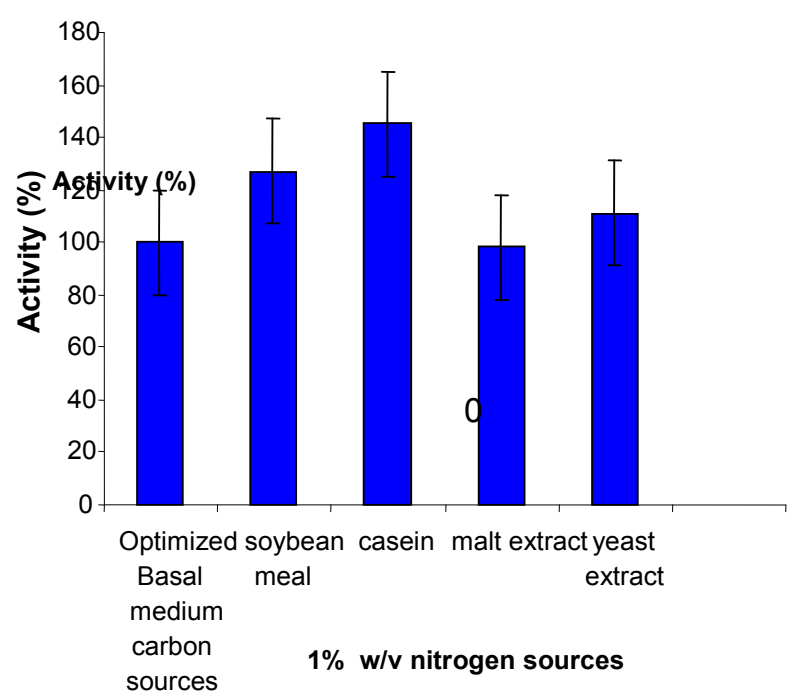

Figure.4 Effect of nitrogen sources on protease production

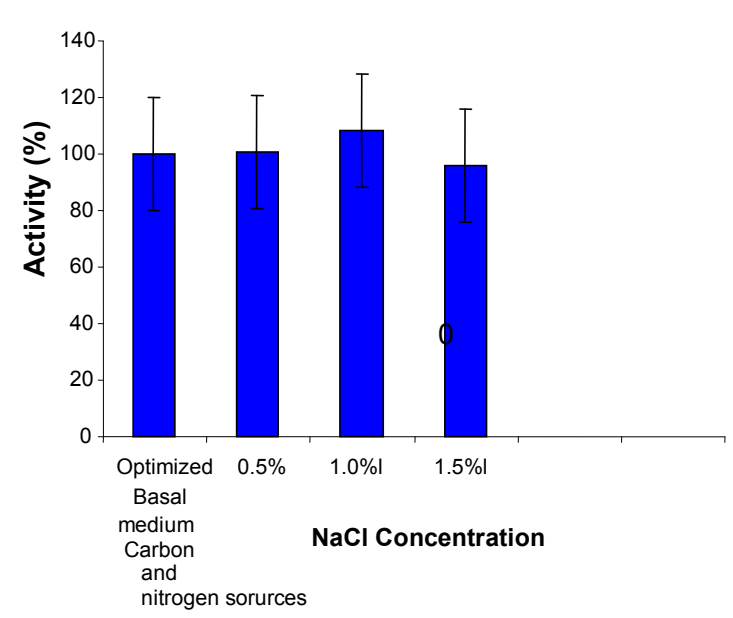

Figure 5 Effect of NaCl on protease production

## 3. CONCLUSION

The isolate having higher protease activity was selected for biochemical characterization and identification. The organism was identified as Streptomyces. Further, authentic identification at IMTECH, Chandigarh reveled that organism belongs to *Streptomyces pulveraceus*. This is the first report on alkaline protease production by (*Streptomyces pulveraceus* MTCC 8374). The optimized conditions of the fermentation media can be implemented in large scale for production of alkaline proteases.